\documentclass[a4paper]{jpconf}
\usepackage{amssymb,latexsym,amsmath}
\usepackage[dvips]{graphicx}
\usepackage{cite}
\newtheorem{theo}{Theorem}
\newtheorem{defi}[theo]{Definition}

\newtheorem{prop}[theo]{Proposition}
\def\mi{\mathrm{i}} 

\def\nn{\nonumber}

\def\C{{\cal C}}
\def\P{{\cal P}}
\def\su{\mathfrak{su}}

\def\sl{\mathfrak{sl}}

\def\diag{\mathop{\rm diag}\nolimits}
\def\qdots{\mathinner{\mkern1mu\raise1pt\vbox{\kern7pt\hbox{.}}\mkern2mu
 \raise4pt\hbox{.}\mkern2mu\raise7pt\hbox{.}\mkern1mu}}
\newcommand{\myatop}[2]{\genfrac{}{}{0pt}{}{#1}{#2}}
\def\mybox{\hfill$\Box$}
\begin{document}
\title{The $\su(2)$ Krawtchouk oscillator model under the $\C\P$ deformed symmetry}
\author{E I Jafarov$^1$, A M Jafarova$^2$, and J Van der Jeugt$^3$}

\address{$^1$ Institute of Physics, Azerbaijan National Academy of Sciences, Javid av. 33, AZ-1143 Baku, Azerbaijan}
\address{$^2$Institute of Mathematics and Mechanics, Azerbaijan National Academy of Sciences, Baku, Azerbaijan}
\address{$^3$Department of Applied Mathematics, Computer Science and Statistics, Ghent University, Krijgslaan 281-S9, B-9000 Gent, Belgium}
\ead{ejafarov@physics.ab.az; Joris.VanderJeugt@UGent.be}


\begin{abstract}
We define a new algebra, which can formally be considered as a $\C\P$ deformed $\su(2)$ Lie algebra. 
Then, we present a one-dimensional quantum oscillator model, of which the wavefunctions of even and odd states are expressed by Krawtchouk polynomials with fixed $p=1/2$, $K_{2n}(k;1/2,2j)$ and $K_{2n}(k-1;1/2,2j-2)$. 
The dynamical symmetry of the model is the newly introduced $\su(2)_{\C\P}$ algebra. 
The model itself gives rise to a finite and discrete spectrum for all physical operators (such as position and momentum). 
Among the set of finite oscillator models it is unique in the sense that any specific limit reducing it to a known oscillator models does not exist.
\end{abstract}

\section{Introduction}

Krawtchouk polynomials are the simplest finite-discrete polynomials of the Askey scheme of orthogonal polynomials.
They play an essential role in the study of problems coming from different branches of modern science. 
Image analysis based on their moments~\cite{yap2003}, exactly solvable birth and death processes~\cite{sasaki2009}, perfect qubit transfer in a spin chain with nearest-neighbour interaction~\cite{albanese2004} are only a few of such problems solved by employing Krawtchouk polynomials. 
Also in the ``discretisation'' of the quantum-mechanical formulation of the harmonic oscillator problem they play an important role.
The non-relativistic explicit formulation of this problem in the infinite continuous configuration space is well known within the canonical approach. 
The wavefunctions of stationary states are expressed in terms of the Hermite polynomials both in momentum and position representation~\cite{landau}. 
Under the assumption that the quantum world is quite different from the classical one, various approaches and methods allow to construct a number of exactly solvable mathematical models of the quantum harmonic oscillator, where the wavefunctions are expressed in terms of other known orthogonal polynomials. 
Also the Krawtchouk polynomials allow one to construct a number of interesting quantum harmonic oscillator models in finite-discrete configuration space. 
At present, the known models using Krawtchouk polynomials are the $\su(2)$ model~\cite{Atak1990,Atak2001,Atak2005} and the $\sl(2|1)$ model~\cite{ej} for the finite quantum harmonic oscillator. 
In the case of $\su(2)$ dynamical symmetry, the wavefunctions of the model are defined through the special case of Krawtchouk polynomials with a fixed parameter $p=1/2$.
For the `supersymmetric' model with $\sl(2|1)$ dynamical symmetry, the wavefunctions are expressed by means of general Krawtchouk polynomials with parameter $0<p<1$. 
In the present work, we present a new oscillator model with wavefunctions of even and odd states also expressed by Krawtchouk polynomials with fixed $p=1/2$, but in a very different way: the polynomials that appear are $K_{2n}(k;1/2,2j)$ and $K_{2n}(k-1;1/2,2j-2)$ respectively. 
The dynamical symmetry is formally called a $\C\P$ deformation of the $\su(2)$ Lie algebra.

The current paper is structured as follows; in section~2, the $\C\P$ deformation of the $\su(2)$ Lie algebra and its representations are introduced. 
Then, this algebra is used for the construction of a model of the finite-discrete quantum harmonic oscillator. 
In section~4, we discuss explicit expression of the wavefunctions of the constructed model both in momentum and position representations. 
Finally, section~5 covers a brief summary and discussion of the obtained results.

\section{The algebra $\su(2)$ under the $\C\P$ deformed symmetry}

Our starting point is the definition of the formal algebra $\su(2)$ under the $\C\P$ deformed symmetry by its basis elements $J_0$, $J_+$ and $J_-$. 

\begin{defi}
The algebra $\su(2)_{\C\P}$ is a unital algebra with basis elements $J_0$, $J_+$, $J_-$, $\C$ and $\P$ subject to 
the following relations:
\begin{itemize}
\item The operator $\C$ commutes with all basis elements.
\item $\P$ is a parity operator satisfying $\P^2=1$ and
\begin{equation}
[\P,J_0]=\P J_0-J_0\P=0, \qquad \{\P,J_\pm\}=\P J_\pm + J_\pm \P= 0.
\label{P}
\end{equation}
\item The $\su(2)$ commutation relations are $\C\P$ deformed as follows:
\begin{align}
& [J_0, J_\pm] = \pm J_\pm,  \label{J0J+} \\
& [J_+, J_-] = 2J_0 \left( {\C\P-1} \right).
\label{J+J-}
\end{align}
\end{itemize}
\end{defi}

The next step is to describe representations of this algebra, by defining the actions of the basis elements $J_0$, $J_+$ and $J_-$ in such a way that they are compatible with the commutation relations \eqref{J0J+} and \eqref{J+J-}. 
It is easy to verify that an action exists on the usual standard basis vectors $\left| {j,m} \right\rangle $ (with $j$ being an integer and $m =  - j, - j + 1, \ldots , + j$), which is reminiscent of but different from the common $\su(2)$ algebra action. This is given by:
\begin{align}
& J_0 |j,m\rangle = m\;|j,m\rangle, \label{act-J0}\\
& J_+ |j,m\rangle = 
  \begin{cases}
 \sqrt{(j-m)(j-m - 1)}\;|j,m+1\rangle, & \text{if $j+m$ is even;} \\
 \sqrt{(j+m)(j+m+1)}\;|j,m+1\rangle, & \text{if $j+m$ is odd,}  
 \end{cases}\label{act-J+}\\
& J_- |j,m\rangle = 
  \begin{cases}
 \sqrt{(j+m)(j+m-1)}\;|j,m-1\rangle, & \text{if $j+m$ is even;} \\
 \sqrt{(j-m)(j-m+1)}\;|j,m-1\rangle, & \text{if $j+m$ is odd.} 
 \end{cases} \label{act-J-}
\end{align} 
Note that $j$ must be integer (and not half-integer) in order to be a representation (one way to see it is that one should still have $J_+|j,j\rangle =0$ and $J_-|j,-j\rangle =0$). The algebra representation is completed by the action of the extra operators $\C$ and $\P$:
\begin{align}
& \C |j,m\rangle = 2j\;|j,m\rangle, \label{act-C}\\
& \P |j,m\rangle = (-1)^{j+m}\;|j,m\rangle. \label{act-P}
\end{align} 

Despite the fact that this algebra is at first sight very close to $\su(2)$, its behaviour is quite different.
For example, the representations matrices of $J_0, J_{\pm}$ in the $\su(2)_{\C\P}$ algebra never coincide with the representation matrices of these elements in the $\su(2)$ algebra, except for the trivial representation $j=0$.
Note also that there is no ``deformation parameter'' that takes the $\su(2)_{\C\P}$ algebra to the $\su(2)$ algebra for some limit.

\section{A one-dimensional oscillator model based on the algebra $\su(2)_{\C\P}$}

We now consider an oscillator model based on the algebra introduced, and choose the Hamiltonian, position and momentum operators in a way that is similar to the non-deformed $\su(2)$ one-dimensional oscillator case~\cite{Atak2005,Atak2001}, i.e.
\begin{equation}
\hat H = J_0+j+\frac12, \qquad 
\hat q = \frac12 (J_++J_-), \qquad
\hat p = \frac{\mi}{2}(J_+-J_-).
\end{equation}
One can easily check that these three operators satisfy the Heisenberg equations:
\begin{equation}
\label{heisenberg}
\left[ {\hat H,\hat q} \right] =  - \mi\hat p,\quad \left[ {\hat H,\hat p} \right] = \mi\hat q.
\end{equation}
Then, one can also observe that 
\begin{equation}
\hat H |j,m\rangle = (m+j+\frac12)|j,m\rangle.
\end{equation}
So the spectrum of $\hat H$ is equidistant but finite:
\begin{equation}
n+\frac12 \qquad (n=0,1,\ldots,2j).
\end{equation}
From the actions~\eqref{act-J+} and~\eqref{act-J-} one obtains for the position operator $\hat q$ that
\begin{equation}
\label{krawt-ff-1}
2\hat q\left| {j,m} \right\rangle  = \sqrt {\left( {j + m} \right)\left( {j + m - 1} \right)} \left| {j,m - 1} \right\rangle  + \sqrt {\left( {j - m} \right)\left( {j - m - 1} \right)} \left| {j,m + 1} \right\rangle,
\end{equation}
when $j+m=2n$ is even, and
\begin{equation}
\label{krawt-ff-2}
2\hat q\left| {j,m} \right\rangle  = \sqrt {\left( {j - m} \right)\left( {j - m + 1} \right)} \left| {j,m - 1} \right\rangle  + \sqrt {\left( {j + m} \right)\left( {j + m + 1} \right)} \left| {j,m + 1} \right\rangle ,
\end{equation}
when $j+m=2n+1$ is odd. 
One can do the same computations for the operator $2\mi\hat p$.
In matrix form, both position and momentum operators $2\hat q$ and $2\mi \hat p$ take the following form:
\begin{equation}
\label{m-q}
2\hat q = \left( {\begin{array}{*{20}c}
   0 & {M_0 } & 0 &  \cdots  & 0  \\
   {M_0 } & 0 & {M_1 } &  \cdots  & 0  \\
   0 & {M_1 } & 0 &  \ddots  & {}  \\
    \vdots  &  \vdots  &  \ddots  &  \ddots  & {M_{2j - 1} }  \\
   0 & 0 & {} & {M_{2j - 1} } & 0  \\
\end{array}} \right) \equiv M^q,
\end{equation}
and
\begin{equation}
\label{m-p}
2\mi \hat p = \left( {\begin{array}{*{20}c}
   0 & {M_0 } & 0 &  \cdots  & 0  \\
   { - M_0 } & 0 & {M_1 } &  \cdots  & 0  \\
   0 & { - M_1 } & 0 &  \ddots  & {}  \\
    \vdots  &  \vdots  &  \ddots  &  \ddots  & {M_{2j - 1} }  \\
   0 & 0 & {} & { - M_{2j - 1} } & 0  \\
\end{array}} \right) \equiv M^p,
\end{equation}
with matrix elements
\begin{equation}
\label{m-k}
M_k  = 
  \begin{cases}
 \sqrt {k\left( {k + 1} \right)}, & \text{if $k$ is odd;} \\
 \sqrt {\left( {2j - k} \right)\left( {2j - k - 1} \right)}, & \text{if $k$ is even.}
 \end{cases}
\end{equation}

The purpose is now to study the eigenvalues of the position (and momentum) operator in this model.
In other words: we need to find eigenvalues and eigenvectors of these matrices $M^q$ and $M^p$.
It turns out that these eigenvectors can be constructed in terms of symmetric Krawtchouk polynomials. 
In general, Krawtchouk polynomials $K_n \left( {x;p,N} \right)$ of degree $n$ ($n=0,1,\ldots,N$) in the variable $x$, 
with parameter $0<p<1$, are defined in terms of the $_2 F_1$ hypergeometric function of argument $\frac{1}{p}$ as follows~\cite{kls}:
\begin{equation}
\label{kr-def}
K_n \left( {x;p,N} \right) = {\ }_2F_1 \left( \myatop{-n,-x}{-N}; \frac{1}{p} \right).
\end{equation}
Their orthogonality holds for discrete values of $x$, and is given by:
\begin{equation}
\label{kr-orth}
\sum_{x = 0}^N {w\left( {x,N} \right)K_n \left( {x;p,N} \right)K_{n'} \left( {x;p,N} \right)}  = h\left( {n,N} \right)\delta _{n,n'} ,
\end{equation}
where,
\begin{align}
\label{kr-w}
& w\left( {x,N} \right) = \binom{N}{x} p^x (1-p)^x \quad (x = 0,1, \ldots ,N ), \\ 
\label{kr-h}
& h\left( {n,N} \right) = \frac{{\left( { - 1} \right)^n n!}}{{\left( { - N} \right)_n }}\left( {\frac{{1 - p}}{p}} \right)^n .
\end{align}
Here, $(a)_k=a(a+1)\cdots(a+k-1)$ are Pochhammer symbols. 
We can introduce the following orthonormal Krawtchouk functions:
\begin{equation}
\label{kr-orthon}
\tilde K_n \left( {x;p,N} \right) = \sqrt {\frac{{w\left( {x,N} \right)}}{{h\left( {n,N} \right)}}} K_n \left( {x;p,N} \right).
\end{equation}
Let us also mention that for $p=1/2$, the corresponding Krawtchouk polynomials are called ``symmetric'', since
\begin{equation}
K_n(x;\textstyle{\frac12},N) = (-1)^n K_n(N-x;\textstyle{\frac12},N).
\end{equation}

In previous examples, the construction of a quantum oscillator model in term of orthogonal polynomials is based on the differential (or difference) equation, the solution of which is expressed in terms of that polynomial. 
In this sense, the $\su(2)$ one-dimensional Krawtchouk oscillator model~\cite{Atak2005,Atak2001} is based on the difference equation for Krawtchouk polynomials~\cite[(9.11.5)]{kls}, whereas the $\sl(2|1)$ one-dimensional Krawtchouk oscillator model~\cite{ej} is based on the forward and backward shift operator equations for them~\cite[(9.11.6), (9.11.8)]{kls}. 
For our current construction, the essential point here is a set of new difference equations for symmetric Krawtchouk polynomials ($p=\frac{1}{2}$). 
These new relations involve a pair of symmetric Krawtchouk polynomials shifted in variables $2x$ or $2(x + 1)$ and parameters $2(j-1)$ and $2j$.

\begin{prop}
The symmetric Krawtchouk polynomials satisfy the following difference equations:
\begin{align}
& j\left( {2j - 1} \right)K_{j + n} \left( {2\left( {x + 1} \right);\textstyle{\frac12},2j} \right) =  - \left( {x + 1} \right)\left( {2x + 1} \right)K_{j + n - 1} \left( {2x;\textstyle{\frac12},2\left( {j - 1} \right)} \right) \nonumber \\ 
& \qquad + \left( {j - x - 1} \right)\left( {2j - 2x - 3} \right)K_{j + n - 1} \left( {2\left( {x + 1} \right);\textstyle{\frac12},2\left( {j - 1} \right)} \right); \label{kr-1}
\\ 
& 2\left( {j + n} \right)\left( {j - n} \right)K_{j + n - 1} \left( {2x;\textstyle{\frac12},2\left( {j - 1} \right)} \right) = j\left( {2j - 1} \right)K_{j + n} \left( {2x;\textstyle{\frac12},2j} \right) \nonumber\\ 
& \qquad - j\left( {2j - 1} \right)K_{j + n} \left( {2\left( {x + 1} \right);\textstyle{\frac12},2j} \right). \label{kr-2}
\end{align}
In these equations, the integer $n$ is arbitrary but should be taken appropriately (e.g.\ for the first Krawtchouk polynomial, the degree $j+n$ should belong to
$\{ 0,1,\ldots,2j\}$).
\end{prop}

\noindent {\bf Proof.}
The validity of both equations is related to a higher level hypergeometric series than ${}_2F_1$, namely ${}_3F_2$.
For this purpose, we can use the following relation~\cite[(39)]{esj}:
\begin{equation}
\label{3f2-2f1}
{}_2F_1 \left( \myatop{-2x, -j-n}{-2j};2 \right) = (-1)^x \frac{\binom{j}{x}}{\binom{2j}{2x}}\ 
{}_3F_2 \left( \myatop{-n,n,-x}{1/2,-j}; 1 \right).
\end{equation}
Using this equation for the ${}_2F_1$ expressions in~\eqref{kr-1} leads to the following identity to verify:
\begin{align}
& j\; {}_3 F_2 \left( \myatop{- n,n, - x - 1}{1/2, - j} ; 1\right) 
= (x + 1) \;{}_3F_2 \left( \myatop{ - n,n, - x}{1/2, - j + 1} ;1 \right) \nn\\
&\qquad - (x - j + 1)\;{}_3F_2 \left( \myatop{ - n,n, - x - 1}{1/2, - j + 1} ;  1 \right). \label{kr-11}
\end{align}
Now, to prove the correctness of~\eqref{kr-11} it is sufficient to simplify the Pochhammer symbols in the expansion of the ${}_3F_2$'s in the right hand side, collect terms, and the left hand side appears automatically.

The proof of~\eqref{kr-2} is similar. In this case, using~\eqref{3f2-2f1} leads to the following identity:
\begin{align}
&\frac{{j^2  - n^2 }}{j} \;{}_3F_2 \left( \myatop{ - n,n, - x}{1/2, - j + 1} ;1 \right) 
= ( x + \frac12) \;{}_3F_2 \left( \myatop{ - n,n, - x - 1}{1/2, - j} ; 1 \right) \nonumber\\
&\qquad - ( x - j + \frac12 )\;{}_3F_2 \left( \myatop{ - n,n, - x}{1/2, - j} ;1 \right).\label{kr-22}
\end{align}
The verification of this again depends on elementary manipulations of Pochhammer symbols.
\mybox

Now, our aim is to construct eigenvalues and eigenvectors of $M^q$~\eqref{m-q}.
This is rather technical, but is essentially based on~\eqref{kr-1} and~\eqref{kr-2}.
For this purpose, we construct the matrix $U$ in the following proposition.

\begin{prop}
Let $M^q\equiv 2\hat q$ be the tridiagonal $(2j+1)\times(2j+1)$-matrix~\eqref{m-q} and let $U=(U_{kl})_{0\leq k,l\leq 2j}$
be the $(2j+1)\times(2j+1)$-matrix with matrix elements as follows.
For an even row index $2i$ ($i=0,1,\ldots,j$), let
\begin{align}
& U_{2i,j-k} = U_{2i,j+k} = (-1)^i 2^{k-j} \tilde K_{2i}(k;\textstyle{\frac{1}{2}},2j), \; k\in\{1,\ldots,j-1\};  \label{Ueven}\\
& U_{2i,0} = U_{2i,2j} = (-1)^i 2^{-1/2} \tilde K_{2i}(j;\textstyle{\frac{1}{2}},2j); \nonumber\\
& U_{2i,j} =(-1)^i 2^{-j+1/2} \tilde K_{2i}(0;\textstyle{\frac{1}{2}},2j).\nonumber
\end{align}
For an odd row index $2i+1$ ($i=0,1,\ldots,j-1$), let
\begin{align}
& U_{2i+1,j-k} = -U_{2i+1,j+k} = (-1)^{i+1} 2^{k-j} \tilde K_{2i}(k-1;\textstyle{\frac{1}{2}},2j-2), \; k \in\{1,\ldots,j-1\}\label{Uodd}\\
& U_{2i+1,0} = -U_{2i+1,2j} = (-1)^{i+1} 2^{-1/2} \tilde K_{2i}(j-1;\textstyle{\frac{1}{2}},2j-2); \nn\\
& U_{2i+1,j} = 0. \nn
\end{align}
Then $U$ is an orthogonal matrix:
\begin{equation}
U U^T = U^TU=I.
\end{equation}
Furthermore, the columns of $U$ are the eigenvectors of $M^q$, i.e.
\begin{equation}
M^q U = U D^q,
\label{MUUD}
\end{equation}
where $D^q= \diag (\epsilon_0,\epsilon_1,\ldots,\epsilon_{2j})$ is a diagonal matrix 
containing the eigenvalues $\epsilon_k$ of $M^q$:
\begin{equation}
\epsilon_{k}=-2\sqrt{(j-k)(j +k)}, \quad \epsilon_{j}=0, \quad \epsilon_{2j-k}=2\sqrt{(j-k)(j+k)}, \label{epsilon} 
\quad (k=0,\ldots,j-1).
\end{equation}
\label{propU}
\end{prop}

\noindent {\bf Proof.}
The proof of this proposition is quite similar to~\cite[Proposition~3]{esj}. 
The matrix relations~\eqref{MUUD} follow directly from the two difference equations~\eqref{kr-1} and \eqref{kr-2}.
The orthogonality of the matrix $U$ is slightly tricky, and does not follow (as one would expect) directly from the orthogonality of symmetric Krawtchouk polynomials.
Instead, just as in the proof of Proposition~2 we need to go to higher level hypergeometric series.
Reexpressing the ${}_2F_1$'s in terms of ${}_3F_2$'s by means of~\eqref{3f2-2f1}, the corresponding ${}_3F_2$'s can be related to Hahn polynomials~\cite{kls} with parameters $\alpha=\beta=-1/2$.
The orthogonality of these Hahn polynomials leads to:
\begin{equation}
\sum_{x=0}^N \frac{1}{2^{2N-1}} \binom{2N}{n} \binom{2N}{2x} K_{n}(2x;\textstyle{\frac12},2N) K_{n'}(2x;\textstyle{\frac12},2N) = 
\left\{ \begin{array}{ll}
   \delta_{n,n'}, & \hbox{if } n\ne N,\\[2mm]
  2\delta_{n,n'}, & \hbox{if } n=N,
  \end{array} \right.
\label{Korth}  
\end{equation}
for $n,n'\in\{0,1,\ldots,N\}$.
The extra factor 2 for $n=n'=N$ appears because for that case one has to take a limit of the squared norm of the Hahn polynomial, so one gets another factor than in the other cases where one can just substitute $\alpha=\beta=-1/2$. 
Equation~\eqref{Korth} is the underlying equation proving the orthogonality of the columns of $U$.
The extra factor 2 is also responsible for the extra factors $2^{1/2}$ when $k=j$.
\mybox

This proposition gives us essentially the eigenvalues $q$ of the position operator $\hat q$. 
There are $2j+1$ distinct eigenvalues and they are given by:
\begin{equation}
\label{pos-spec}
 - j, - \sqrt {j^2  - 1} , \ldots  - \sqrt {j^2  - \left( {j - 1} \right)^2 } ,0,\sqrt {j^2  - \left( {j - 1} \right)^2 } , \ldots ,\sqrt {j^2  - 1} ,j,
\end{equation}
or in general:
\begin{equation}
\label{q-k}
q_{ \pm (j - k)}  =  \pm \sqrt {\left( {j - k} \right)\left( {j + k} \right)} ,\quad k = 0,1,2, \ldots ,j.
\end{equation}

In Figure~1, we present a plot of the spectrum of the position operator for the $\su(2)$, $\sl(2|1)$ and $\su(2)_{\C\P}$ oscillator models. 
One observes that the spectrum of the position operator is quite different compared to the finite-discrete oscillator models already known. 
The main difference is the distribution of positions. 
For the $\su(2)$ oscillator model, the positions are equidistant ($q_{\pm k}^{\su(2)}=\pm k$).
The distribution of positions for the $\sl(2|1)$ model is not equidistant, due to $q_{\pm k}^{\sl(2|1)}=\pm \sqrt{k}$. 
In that case, the interval covering the positions is smaller than the corresponding one for the $\su(2)$ case. 
Comparing the $\su(2)_{\C\P}$ case with two others, we observe that some of its properties overlap with $\su(2)$ and others with $\sl(2|1)$. 
The interval covering the positions is the same for the $\su(2)$ and the $\su(2)_{\C\P}$ model: $q_{\pm j}^{\su(2)}=q_{\pm j}^{\su(2)_{\C\P}}$. 
Both the $\sl(2|1)$ and $\su(2)_{\C\P}$ oscillator models have a non-equidistant distribution of the position eigenvalues. 
The main feature of the $\su(2)_{\C\P}$ oscillator model is that the position values are concentrated near the border of the inteval $[-j,+j]$, whereas near the middle of the interval the distribution is sparse.

\begin{figure}[htb]
\begin{center}
\includegraphics[width=0.7\textwidth]{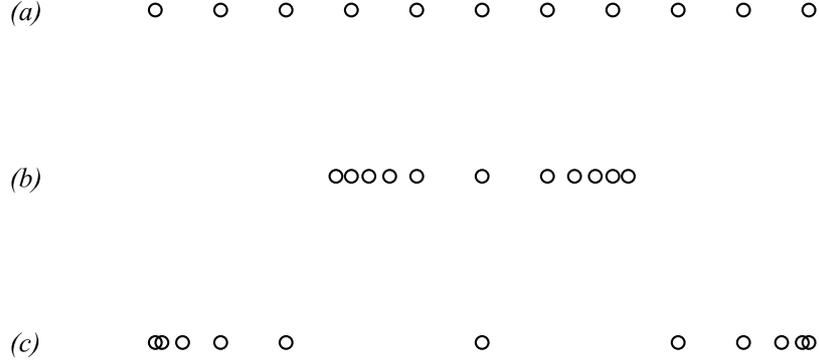}
\end{center}
\caption{Spectrum of the position operator for $j=5$, (a)~in the case of the $\su(2)$ model, (b)~in the case of the $\sl(2|1)$ model and (c)~in the case of the $\su(2)_{\C\P}$ model.}
\end{figure}

\section{Position and momentum wavefunctions}
The position wavefunctions of the $\su(2)_{\C\P}$ finite-discrete oscillator model are defined in a similar way as in~\cite{esj}. 
Their explicit expression is computed through the overlap between the $\hat q$-eigenvectors $\left. {|j,q_{j-k} } \right)$ and the $\hat H$-eigenvectors $|j,m\rangle$.
Let us denote them by $\Psi_{j+m}(q)$, where $m=-j,-j+1,\ldots,+j$, and where $q$ assumes one of the discrete values of $q_k$ with $(k=-j,-j+1,\ldots,+j)$. Simple computations show that
\begin{equation}
\label{wf-pos}
\Psi_{j+m}(q_{j-k})= \langle j,m | j,q_{j-k} ) = U_{j+m,j+k},
\end{equation}
where $U_{j+m,j+k}$ are the matrix elements of $M^q$. 
Then, one can find that for the even case $j+m=2n$ and positive values of position, the wavefunctions have the following expression: 
\begin{equation}
\label{wf-pos-even}
\Psi_{2n} ( q_{k}) = (-1)^n 2^{k-j} \tilde K_{2n}(k;\textstyle{\frac12},2j), \quad n = 0,1, \ldots ,j,\quad k = 1, \ldots ,j - 1,
\end{equation}
whereas, for the odd case $j+m=2n+1$ and positive values of position, we have
\begin{equation}
\label{wf-pos-odd}
\Psi_{2n + 1} ( q_{k}) = (-1)^n 2^{k-j}\tilde K_{2n}(k-1;\textstyle{\frac12},2j-2).
\end{equation}
One can extend these computations and obtain similar expressions for zero and negative values of the positions.

In Figure~2 we plot the discrete wavefunctions~\eqref{wf-pos-even} and~\eqref{wf-pos-odd} for some large $j$-value, and for some $n$-values. 
Due to the unique property of the position eigenvalues to be located near the border of the interval $[-j,+j]$, the behaviour of the wavefunctions also reflects this peculair distrubution.

\begin{figure}[htp]
\begin{center}
\begin{tabular}{cc}
\includegraphics[scale=0.35]{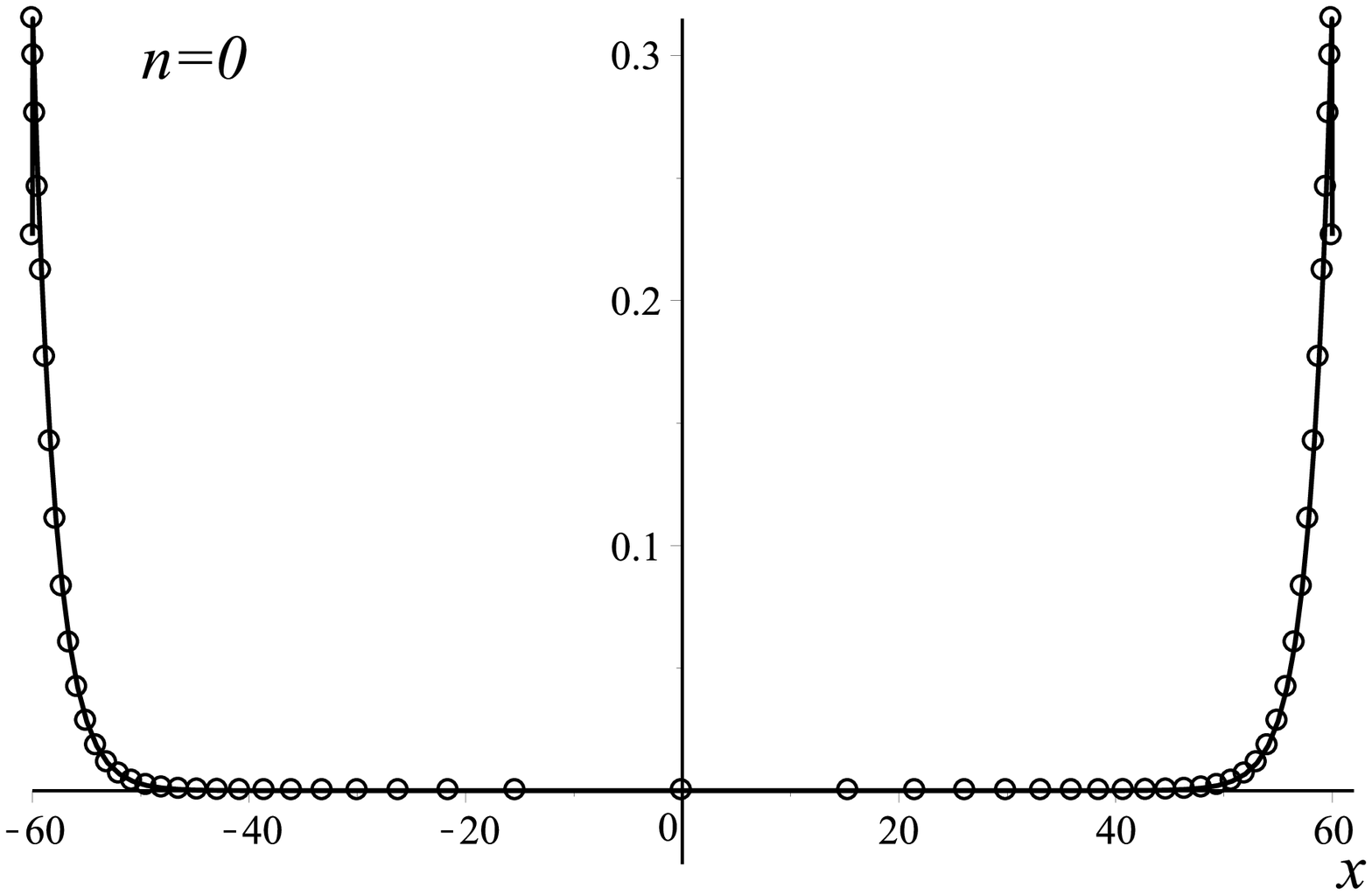} & \includegraphics[scale=0.35]{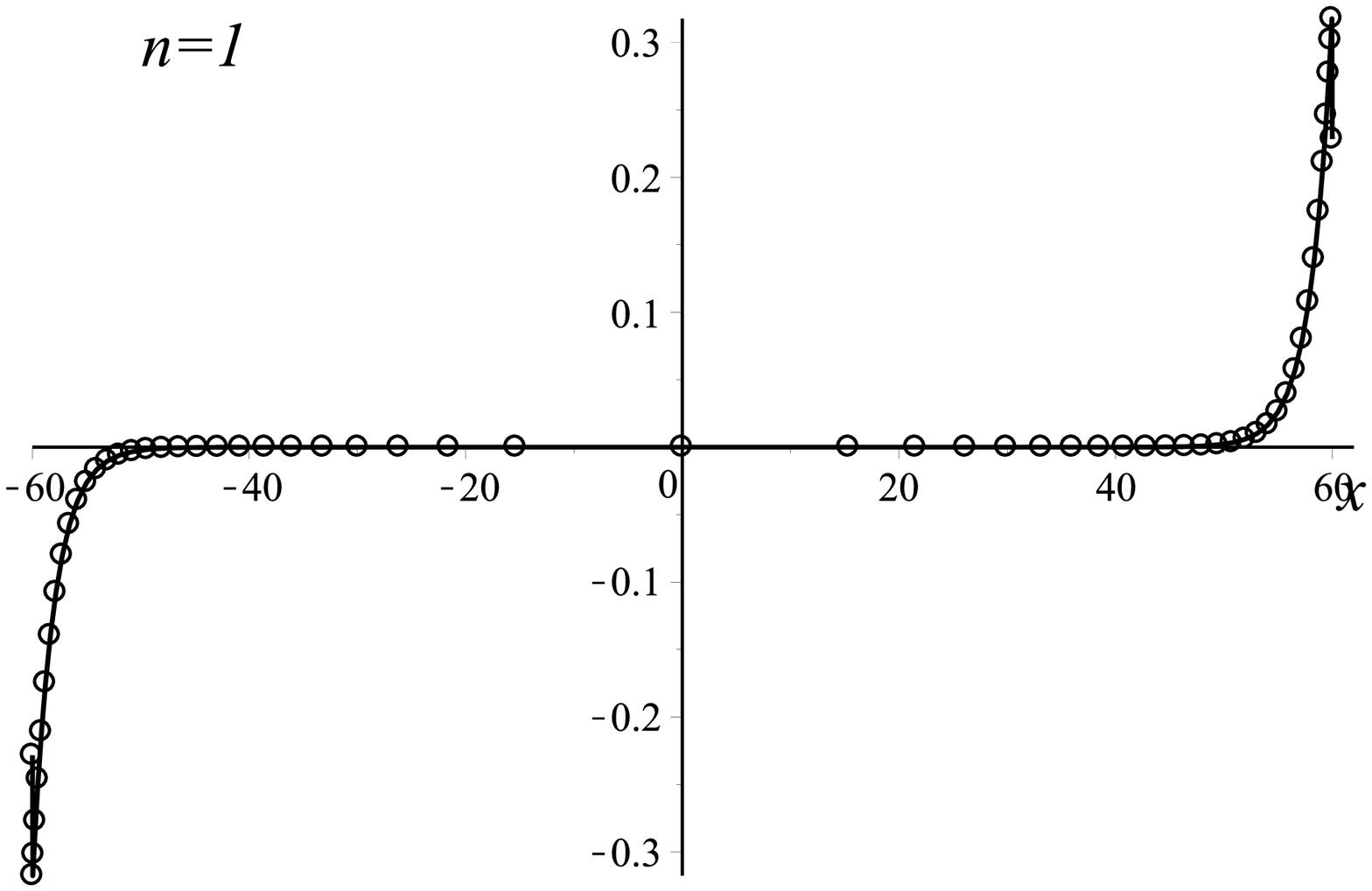} \\
\includegraphics[scale=0.35]{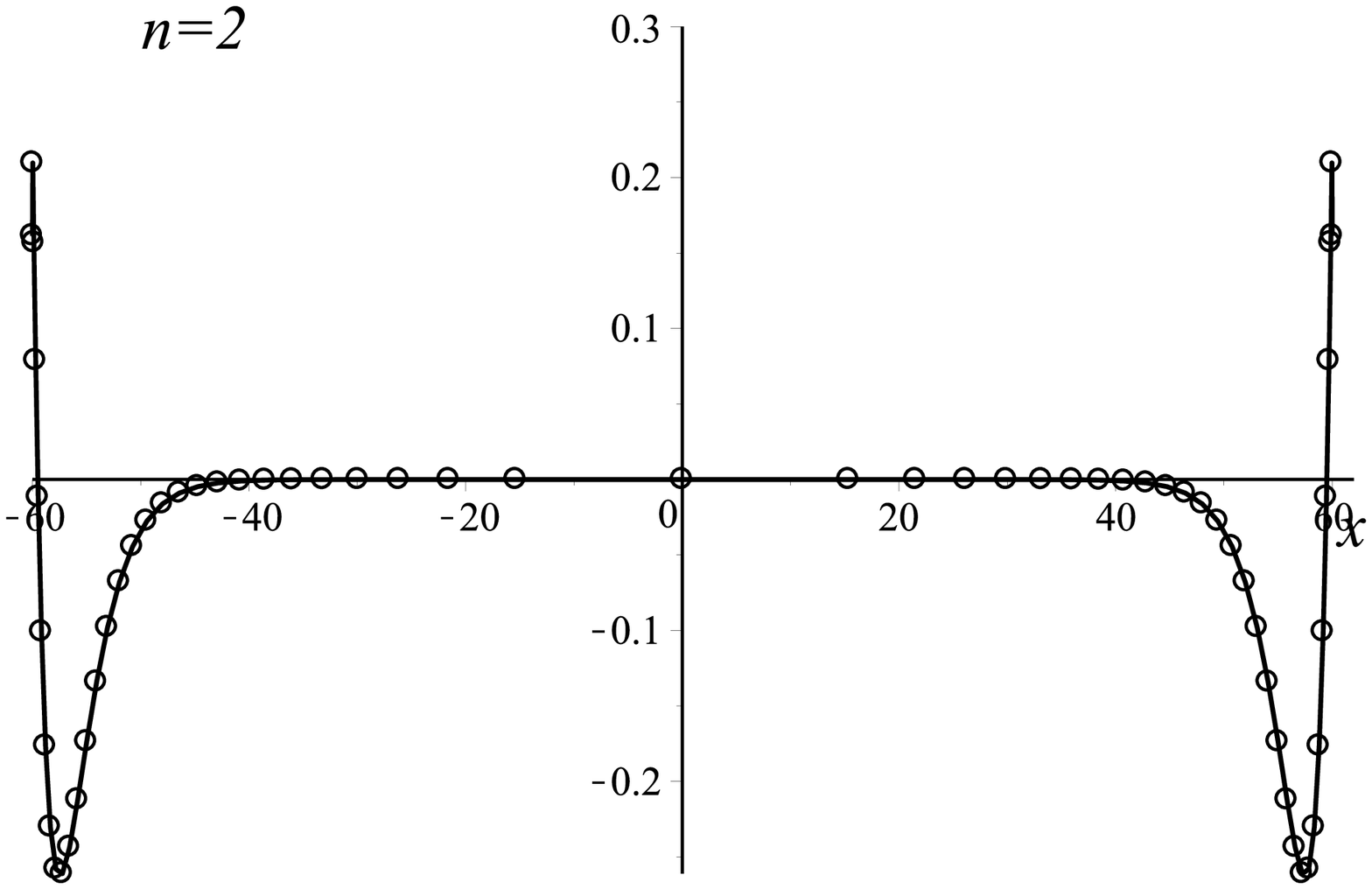} & \includegraphics[scale=0.35]{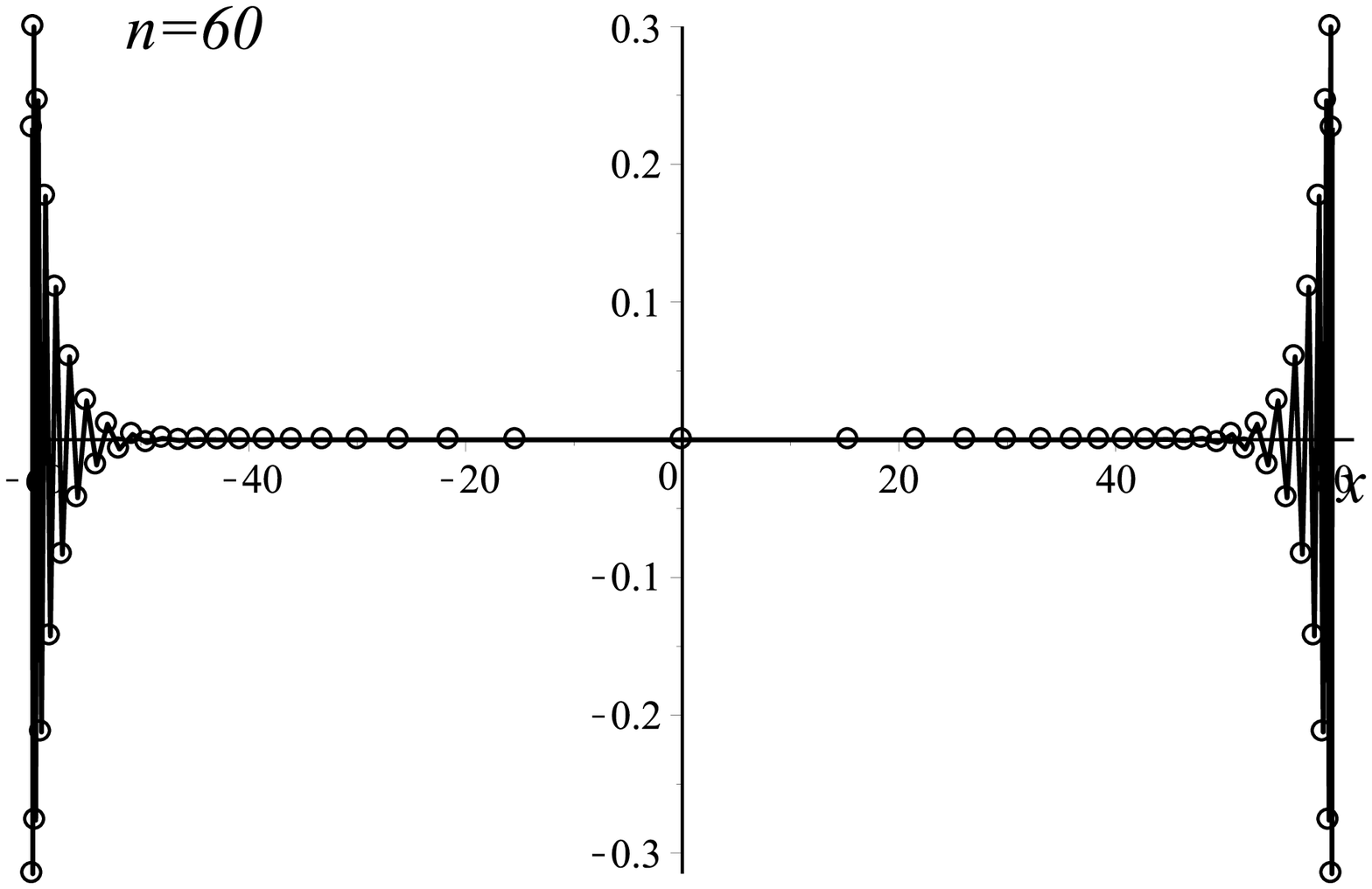} \\[-1mm]
\end{tabular} 
\end{center}
\caption{Plots of the discrete wavefunctions $\Psi_n(q)$ in the representation with $j=30$. 
We plot the wavefunctions for $n=0,1,2,60$.}
\label{fig2}
\end{figure}

We can define momentum wavefunctions $\Phi_{j+m}(p)$ in a similar way. 
Their explicit expression is computed through the overlap between the $\hat p$-eigenvectors $\left. {|j,p_{j-k} } \right)$ and the $\hat H$-eigenvectors $|j,m\rangle$. In this case, the simple relation between $M^p$ and $M^q$ shows that
\begin{equation}
\label{wf-mom}
\Phi_{j+m}(p_{j-k})= \langle j,m | j,p_{j-k} ) = V_{j+m,j+k}, \quad p_{j-k}\equiv q_{j-k}, \quad V_{k,l}\equiv \mi ^{k+1} U_{k,l}.
\end{equation}
Then, in accordance with equations~\eqref{wf-pos-even} and~\eqref{wf-pos-odd}, one finds that for the even case $j+m=2n$ and for positive values of momentum, the wavefunctions have the following expression: 
\begin{equation}
\label{wf-mom-even}
\Phi_{2n} ( p_{k}) = \mi\; 2^{k-j} \tilde K_{2n}(k;\textstyle{\frac12},2j), \quad n = 0,1, \ldots ,j,\quad k = 1, \ldots ,j - 1,
\end{equation}
whereas for the odd case $j+m=2n+1$ and positive values of momentum, we have
\begin{equation}
\label{wf-mom-odd}
\Phi_{2n + 1} ( p_{k}) = - 2^{k-j}\tilde K_{2n}(k-1;\textstyle{\frac12},2j-2).
\end{equation}
These computations can easily be extended for zero and negative values of the momentum. 
Both wavefunctions $\Psi (q)$ and $\Phi(p)$ are related by the so called $\C\P$ symmetric Krawtchouk transform:
\begin{equation}
\label{cp-krawt}
\Phi \left( {p_l } \right) = \sum\limits_{k =  - j}^j {{\cal K}_{kl} \Psi \left( {q_k } \right)} ,
\end{equation}
where, ${\cal K}=U^TV$. The matrix elements ${\cal K}_{k,l}$ can be computed by using~\eqref{Korth}.

\section{Discussion}

Despite the fact that the non-relativistic one-dimensional quantum oscillator in the canonical approach has a nice explicit solution in terms of Hermite polynomials, there are still attempts to construct new oscillator models inspired by different starting points.
One of these starting points is the assumption of discrete space, in which case one is looking for oscillator models with a discrete position spectrum (which could be finite of infinite discrete).
The most popular finite discrete oscillator model~\cite{Atak2005,Atak2001} is based on the $\su(2)$ algebra, and uses Krawtchouk polynomials.

In current paper, we have constructed a new model for the one-dimensional oscillator, with wavefunctions expressed in terms of the Krawtchouk polynomials with parameter $p=1/2$ (the symmetric Krawtchouk polynomials). 
In order to construct a solvable oscillator model, we have first defined a new algebra, formally called the $\su(2)$ algebra under the $\C\P$ deformation.
The representations of this algebra were given and are similar to representations of $\su(2)$, but the action of the raising and lowering operators $J_+$ and $J_-$ differs considerably.
As a consequence, when considering the oscillator model related to this algebra, the matrix representations of position and momentum operators is also different.
In the standard $|j,m\rangle$ bases, the position operator is tridiagonal, and we have shown that its eigenvalues and eigenvectors involve Krawtchouk polynomials.
The matrix relation for the eigenvalue problem is equivalent to certain finite difference equations for Krawtchouk polynomials.

The resulting model is unique in the sense that it does not generalize any other existing oscillator model.
Peculiar properties are the non-equidistant distribution of the position values as well as the behaviour of the wavefunctions around zero and near the border of the maximum and minimum eigenvalues.

\section*{Acknowledgements}
EIJ kindly acknowledges support from Research Grant \textbf{EIF-2012-2(6)-39/08/1} of the Science Development Foundation under the President of the Republic of Azerbaijan.

\end{document}